\documentclass[10pt,journal,compsoc]{IEEEtran}

\usepackage{booktabs}
\usepackage{multirow}
\usepackage{graphicx}
\usepackage{cite}
\usepackage{lipsum}
\usepackage[table,xcdraw]{xcolor}
\usepackage{amssymb}
\usepackage{amsmath,amssymb,amsfonts}
\usepackage{lscape}

\ifCLASSINFOpdf
\else
\fi
\begin{document}
%

\title{Improving Customer Service Chatbots with Attention-based Transfer Learning}

\author{Jordan~J.~Bird
\IEEEcompsocitemizethanks{\IEEEcompsocthanksitem J.J. Bird is with the Computational Intelligence and Applications Research Group (CIA) at Nottingham Trent University, Nottingham, United Kingdom.\protect\\
E-mail: see https://jordanjamesbird.com/contact/}
\thanks{This work has been submitted to the IEEE for possible publication. Copyright may be transferred without notice, after which this version may no longer be accessible.}}

\markboth{Preprint, 2021}%
{Bird: Improving Customer Service Chatbots with Attention-based Transfer Learning}
%



\IEEEtitleabstractindextext{%
\begin{abstract}
With growing societal acceptance and increasing cost efficiency due to mass production, service robots are beginning to cross from the industrial to the social domain. Currently, customer service robots tend to be digital and emulate social interactions through on-screen text, but state-of-the-art research points towards physical robots soon providing customer service in person. This article explores two possibilities. Firstly, whether transfer learning can aid in the improvement of customer service chatbots between business domains. Second, the implementation of a framework for physical robots for in-person interaction. Modelled on social interaction with Twitter customer support accounts, transformer-based chatbot models are initially assigned to learn one domain from an initial random weight distribution. Given shared vocabulary, each model is then tasked with learning another domain by transferring knowledge from the previous. Following studies on 19 different businesses, results show that the majority of models are improved when transferring weights from at least one other domain, in particular those that are more data-scarce than others. General language transfer learning occurs, as well as higher-level transfer of similar domain knowledge, in several cases. The chatbots are finally implemented on Temi and Pepper robots, with feasibility issues encountered and solutions are proposed to overcome them. 
\end{abstract}

}

\maketitle

\IEEEdisplaynontitleabstractindextext

%
\IEEEpeerreviewmaketitle

\IEEEraisesectionheading{\section{Introduction}\label{sec:introduction}}
\IEEEPARstart{T}{he} growing acceptance of autonomous technology within our daily lives seems to have set us on an inevitable path towards society being aided by physical robots in a variety of different situations. The majority of robots that are used today exist within industrial work environments such as manufacturing\cite{esmaeilian2016evolution} and assembly\cite{michalos2010automotive}, as well as exploration of hazardous environments\cite{lunghi2019multimodal}. Given the societal acceptance of robots and their improvement, we can expect to find robots of a more social nature helping in customer service roles in the future. This would not only help the customer, but also the organisation too; in the UK alone, it has been noted that some consumers often wait up to 30 minutes in a queue before being able to speak to a representative\cite{wallop_2021}. Automating some of these processes with NLP systems which can understand issues would reduce a customer's waiting time and reduce pressure on the organisation by either solving the problem and giving advice autonomously or gathering enough useful information during the conversation which can be passed on to a human being who can then solve the issue more efficiently. 

Providing customer service is an important and expensive aspect of business\cite{bygballe2012managing}, often being the largest department in most companies. Many problems are easy to solve, for example, a forgotten password, and yet customer service representatives spend much of their important time on such issues\cite{potter1994new}. Indeed, the conversations between representatives and customers on these issues are unique and nuanced, with exchanges dependent on prior information and vocabulary etc. Based on this, simply regurgitating the same instructions, similarly to that of which a \textit{forgotten password} button on a website's login form may produce, is sub-par compared to customer service experience. Instead, this work proposes to use attention-based chatbots, where models learn to tune attention to prior exchanges in the conversation before producing the next. There is a considerably large amount of conversation data available on social media wherein social exchanges occur between customers and customer service representatives, and thus, this provides us with a useful starting point to train chatbots which perform similar tasks during interaction with human beings. 

The nature of this article is to explore if it is possible to perform transfer learning of language between different domains to improve chatbot models; models which would form part of a framework to provide customer service to human beings through conversation and advice. To give an example, although the problems solved by Amazon and Tesco customer service representatives may differ greatly, there may exist useful knowledge at the lower level, i.e., vocabulary and language, as well as at the higher level, i.e., logical processes and problem solving, which would aid in improving language models if it could be transferred. 

The scientific contributions of this article are as follows:
\begin{enumerate}
    \item Experiments show that deep learning chatbots can be improved by transferring knowledge from other chatbots also trained to provide customer support. 
    \item Low level transfer of knowledge occurs in terms of English language, and higher level transfer learning occurs between domains facing similar customer support requests.
    \item Feasibility studies highlight several difficulties when implementing the chatbots on physical robots, and solutions are proposed to overcome them to enable a Human-Robot Interaction approach to customer support in person. 
\end{enumerate}

The remainder of this article is presented as follows, Section \ref{sec:background} reviews the scientific state-of-the-art of customer service chatbots as well as describing the theory behind attention-based modelling. Following this, Section \ref{sec:method} describes the method followed by the experiments in this study, before the results are presented and discussed in Section \ref{sec:results}. The results section also provides exploration of the models, providing interesting examples of interaction with the chatbots and discussion. Feasibility observations are made and chatbots are implemented on physical robots in Section \ref{sec:robots}. Finally, Section \ref{sec:futurework} describes the future work that could be performed based on the findings of the experiments in this article before concluding the study. 

\section{Background}
\label{sec:background}
Chatbots are autonomous systems which aim to converse with a human user. Falling into two main categories, chatbots can be either open domain (general conversation) or closed domain (aimed at solving a specific task)\cite{lokman2018modern}. Alan Turing's question, ``can machines think?", led to what we now know as the \textit{Imitation Game} or \textit{Turing Test}. This was the proposal that if, under specific conditions, a machine could mimic a human being, then the computer can be said to possess artificial intelligence\cite{turing1950computing}. Turing originally suggested the use of a teleprinter (specific conditions), but modern systems are tested online\cite{floridi2009turing}. 

Customer service chatbots aim to mimic a human being and to help solve customer queries and issues. For example, SuperAgent\cite{cui2017superagent} is a question-answering chatbot that can mine data from web pages and provide information e.g. answering ``does it come with the pen?" with ``yes it does". Deep learning is often leveraged for customer service chatbots in contemporary studies\cite{nuruzzaman2018survey}, such as LSTM sequence-to-sequence learning in \cite{xu2017new} which led to an improvement over information retrieval for social media-based services. Ranoliya et al.\cite{ranoliya2017chatbot} proposed a more classical XML-based approach for University-related queries, achieving impressive results for an automatic question-answering problem in educational support. Often, data received from customers is further analysed with sentiment analysis via either scoring and polarity\cite{feine2019measuring,tran2021exploring} or classification\cite{bird2019high,jia2021chinese}. In this study, the sequence of inputs and attention masking are considered, and so, although not explicitly scoring or classifying sentiment, valence data still exists within the dataset. 

Several methods have been proposed for improving chatbots. For example, recent work on combining reinforcement learning with human-robot interactions showed further improvement by learning from experience\cite{ricciardelli2019self}. Reinforcement learning has also been suggested to improve chatbots with ensemble learning strategies\cite{cuayahuitl2019ensemble}. Ensemble chatbots have been leveraged in studies concerning medicine\cite{bali2019diabot}, mental healthcare\cite{harilal2020caro}, and education\cite{mondal2018chatbot}. In another study, the addition of synthetic data with transformers was shown to improve the ability of chatbots when an attention-based model was used to generate training data to create additional examples \cite{bird2021chatbot}. In a similar line of questioning to this work, the authors in \cite{ilievski2018goal} suggested the possibility of transferring knowledge between domains for chatbot improvement. In the aforementioned, reinforcement learning improved when transferring knowledge between restaurant booking, movie booking, and tourist information. These results are particularly exciting for tourist information since it differs relative to the other two booking domains, and yet improvements were still achieved. Similar goal-oriented experiments were also performed in \cite{hatua2019goal} with attention mechanisms in place of transfer learning as previously described. Transfer learning has also found success in question-answering systems\cite{yu2018modelling}.

Acceptance of physical robots in customer service is growing, with many studies even suggesting that customers prefer robots with simulated emotional feelings\cite{yam2020robots}. Related research has also suggested that humanoid robots are currently the most emotionally acceptable form, with the current state of hyperrealistic robots remaining within the uncanny valley\cite{mori1970bukimi,murphy2017service}. Wirtz et al.\cite{wirtz2018brave} propose that service employees and robots will eventually work in sync, with some tasks dominated by humans and vice versa for robots. Tangible actions were predicted to include autonomous receptionists and porters, and intangible actions were predicted to include information counters, claim processing, chatbots, etc. Indeed, the line between the two is blurred, i.e., a physical receptionist robot, if emulating a human being, would require the additional implementation of a closed domain chatbot. According to Tuomi et al.\cite{tuomi2021applications}, there are many roles that robots can play when providing customer service. These roles are external, internal, and operational. This study focuses on the operational, specifically in support, defined by the aforementioned study as \textit{dealing with routine tasks and freeing human employees to focus on more complex and dynamic situations}. To give a concrete example of how this would be made possible by a chatbot, consider the example given in the introduction; a forgotten password is a relatively easy task that representatives must solve many times per day, and thus this advice could instead be given by an autonomous system, allowing the human operator to instead focus on more difficult tasks that may not be autonomously solved. 

Attention-based modelling and transformers are a relatively new concept in deep learning\cite{vaswani2017attention}. Although the original study was only performed in 2017, there are many examples where the approach has achieved state-of-the-art performance within several areas of NLP such as text synthesis\cite{devlin2018open,radford2019language} and autonomous question answering\cite{shao2019transformer,lukovnikov2019pretrained}, which are of interest to this study in particular. Attention is calculated via a scaled dot product, where attention weights are calculated for each word in the input vector. The weights for the query $W_{q}$, key $W_{k}$, and value $W_{v}$ are calculated as follows:
\begin{equation}
\label{eq:attention}
    Attention(Q,K,V) = softmax \left(  \frac{QK^T}{  \sqrt{d_{k}}  }  \right) V,
\end{equation}
where the query is an object within the sequence (e.g., a word), the keys are vector representations of the input, and values are produced given a query with keys. A transformer calculates multi-headed attention, where a concatenation of multiple $i$ attention heads $h_{i}$ enables a larger network of interconnected attention calculations:
\begin{equation}
\begin{aligned}
    MH(Q,K,V) = Concatenate(h_{1}, ..., h_{h})W^{O} \\
    h_{i} = Attention(QW^{Q}_{i}, KW^{K}_{i}, VW^{V}_{i}),
\end{aligned}
\end{equation}
in terms of multi-headed attention $MH$ of single-attention heads $h$.

\section{Method}
\label{sec:method}
\begin{figure}
    \centering
    \includegraphics[scale=0.8]{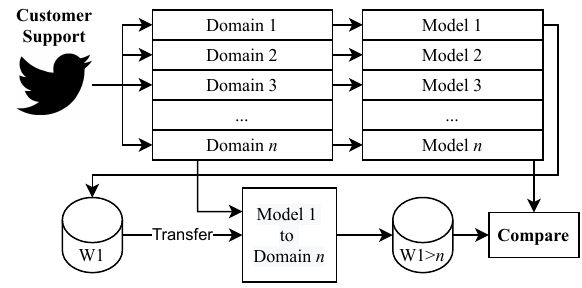}
    \caption{General diagram of the weight transfer learning process between customer support domains. Weights are transferred as a starting distribution before learning to provide support for a different domain.}
    \label{fig:learning-diagram}
\end{figure}
\begin{figure}
    \centering
    \includegraphics[scale=0.8]{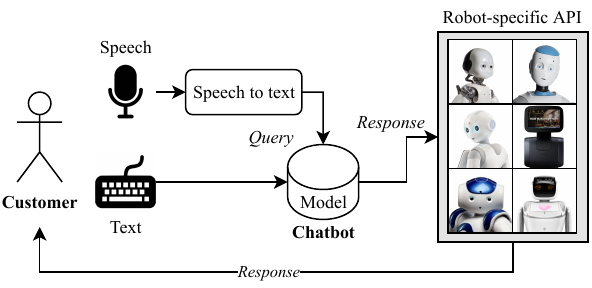}
    \caption{Generalised diagram of the Human-Robot Interaction pipeline. Speech-to-text or text are provided as input to the model derived from the process in Figure \ref{fig:learning-diagram}. Robot-specific APIs are used to respond to the customer. }
    \label{fig:robot-diagram}
\end{figure}
A general diagram of the transfer learning process is shown in Figure \ref{fig:learning-diagram}. A transformer-based chatbot is trained for each domain; then, in the transfer learning experiments, these pretrained weights are used as starting points for learning another domain. Comparing the original and transfer experiments shows the difference between random and other domain weight distributions, thus allowing us to explore whether useful knowledge can be transferred from one domain to another. 
\begin{table}[]
\centering
\caption{Data objects from each domain after pre-processing and cleaning}
\label{tab:dataset}
\begin{tabular}{@{}lll@{}}
\toprule
\textbf{Dataset}                  & \textbf{No. Conversations} & \textbf{File Size (KB)} \\ \midrule
\textit{\textbf{Amazon}}          & 30402                      & 6780                    \\
\textit{\textbf{American Air}}    & 9044                       & 1903                    \\
\textit{\textbf{Apple Support}}   & 29689                      & 6385                    \\
\textit{\textbf{British Airways}} & 6094                       & 1378                    \\
\textit{\textbf{Chipotle}}        & 4350                       & 651                     \\
\textit{\textbf{Comcast}}         & 5581                       & 1246                    \\
\textit{\textbf{Delta}}           & 9163                       & 1784                    \\
\textit{\textbf{Hulu}}            & 5585                       & 1208                    \\
\textit{\textbf{PlayStation}}     & 2783                       & 507                     \\
\textit{\textbf{Sainsburys}}      & 5188                       & 1065                    \\
\textit{\textbf{Spectrum}}        & 4749                       & 1045                    \\
\textit{\textbf{Spotify}}         & 8696                       & 1753                    \\
\textit{\textbf{Sprint}}          & 3054                       & 620                     \\
\textit{\textbf{Tesco}}           & 9273                       & 2180                    \\
\textit{\textbf{TMobile}}         & 6835                       & 1429                    \\
\textit{\textbf{Uber}}            & 9166                       & 1924                    \\
\textit{\textbf{UPS}}             & 3182                       & 782                     \\
\textit{\textbf{Virgin Trains}}   & 5912                       & 1037                    \\
\textit{\textbf{Xbox}}            & 5041                       & 1007                    \\ \bottomrule
\end{tabular}
\end{table}
A dataset of Tweets is initially retrieved from \cite{vector_2017}. The dataset contains 3,003,124 instances of Tweets and responses to and from 19 different support accounts. Tweets that are not written in English are removed. All text is converted to the lower case, and then the human names and punctuation are removed from the strings. To keep profane words from the vocabulary of the chatbot, a list of words banned from Google autocomplete are retrieved from \cite{gabriel_2016}, and any Tweets containing one or more of these terms are removed. The details of each domain post-processing can be observed in Table \ref{tab:dataset}. Across all Twitter conversations, there are 91967 unique tokens that constitute the universal vocabulary. Sparse categorical cross-entropy $-\sum_{c=1}^My_{o,c}\log(p_{o,c})$ is used due to the large number of classes. Due to this, sparse metrics are measured in terms of loss and accuracy. Additionally, top-5 and top-10 accuracy metrics are also considered. 

A total of 361 chatbots are trained, 19 are trained classically from their respective datasets and 342 are trained via transfer learning (from the prior 19 to all of the others). Observations showed that training would take approximately ten days for data split validation models in terms of computational time. For this reason, data splitting (70:30) is chosen as the validation approach given that k-fold and leave-one-out strategies would take considerably more time, rendering the experiments impossible to perform. 

All models in this article were implemented with TensorFlow, executed on a server with shared resources. The model had access to an Intel Xeon E5-2640 v4 CPU (2.4GHz) and performed operations via CUDA on a single Nvidia Tesla M60 Accelerator (with two GPUs on board); the Accelerator operated with 4096 CUDA cores and 16GB GDDR5 VRAM. For privacy reasons, it could not be observed whether there were other users making use of the shared computational resources throughout.

\section{Results and Implementation}
\label{sec:results}
\begin{table}[]
\centering
\caption{Validation loss during the tuning of network topologies (10 epochs on all data) for the selection of attention heads and feed-forward neurons.}
\label{tab:topology}
\begin{tabular}{@{}llllll@{}}
\toprule
\multicolumn{2}{l}{}                                                                                                & \multicolumn{4}{l}{\textbf{Attention Heads}}                                           \\ \midrule
\multirow{5}{*}{\textit{\textbf{\begin{tabular}[c]{@{}l@{}}Dense \\ Neurons\end{tabular}}}} &                       & \textit{\textbf{2}} & \textit{\textbf{4}} & \textit{\textbf{8}} & \textit{\textbf{16}} \\
                                                                                            & \textit{\textbf{64}}  & 2.67                & 2.65                & 2.66                & 2.67                \\
                                                                                            & \textit{\textbf{128}} & 2.68               & 2.65               & 2.65               & 2.67               \\
                                                                                            & \textit{\textbf{256}} & 2.67               & 2.65               & 2.64               & 2.65               \\
                                                                                            & \textit{\textbf{512}} & 2.69               & 2.66              & 2.65               & 2.65               \\ \bottomrule
\end{tabular}
\end{table}
Initially, the network topology is explored. Table \ref{tab:topology} shows the results for tuning the 16 different topologies, where a batch search of $\{2,4,8,16\}$ attention heads and $\{64,128,256,512\}$ neurons in the dense layer are benchmarked. The validation results are relatively marginal following the 10-epoch training exercise on all data, and the lowest overall was that of 8 attention heads with 256 neurons (2.64). Although the differences in the results are small, this topology is chosen for the remaining chatbot experiments for simplicity.

\begin{table}[]
\centering
\caption{Exploration of the effects of reducing vocabulary size (10 epochs on all data).}
\label{tab:vocab-table}
\begin{tabular}{@{}lll@{}}
\toprule
\textbf{Vocabulary Size } & \textbf{Trainable Parameters } & \textbf{Loss } \\ \midrule
\textit{\textbf{10000}}  &  7 386 640                             &   2.51            \\
\textit{\textbf{15000}}  &  9 951 640                             &  2.55             \\
\textit{\textbf{20000}}  &  12 516 640                             &  2.59             \\
\textit{\textbf{25000}}  &  15 081 640                           &   2.63             \\
\textit{\textbf{30000}}  &  17 646 640                            & 2.62   \\ \bottomrule
\end{tabular}
\end{table}
It must be noted that the transformers explored in this work are much smaller than state-of-the-art general transformers (which are large due to the multitude of different tasks they consider). For example, a model made up of 8 attention heads, 256 neurons, and a vocabulary of 30000 has over 17 million trainable parameters as can be observed in Table \ref{tab:vocab-table} compared to GPT-2's 1.5 billion and GPT-3's 175 billion. Table \ref{tab:vocab-table} shows that when adding 20k extra words and therefore over 10 million additional trainable parameters, the preliminary loss metrics do not rise drastically. Given that a difference of +0.11 is observed, a large vocabulary of 30k is chosen for these experiments; this is mainly due to the multiple domains involved to provide ample coverage to each, since the vocabulary is shared between all models to enable direct transfer learning. As described previously, the global vocabulary of all data is made up of 91967 unique tokens, with 30000 being approximately a third of the total tokens present (32.62\%). 

\begin{table}[]
\centering
\caption{Validation metrics after training on domains from an initial random weight distribution.}
\label{tab:no-transfer}
\begin{tabular}{@{}lllll@{}}
\toprule
\textbf{Dataset }                  & \textbf{Loss } & \textbf{Accuracy } & \textbf{Top-5 } & \textbf{Top-10 } \\ \midrule
\textit{\textbf{Amazon}}          &  2.3             &    0.602               &  0.73                       &  0.779                        \\
\textit{\textbf{American Air}}    &  2.17             &    0.645              &   0.754                      &     0.8                     \\
\textit{\textbf{Apple Support}}   &  1.93             &    0.642               &  0.778                       &    0.826                      \\
\textit{\textbf{British Airways}} &   2.19           &   0.658                &   0.764                      &    0.802                      \\
\textit{\textbf{Chipotle}}        &  1.17             &  0.808                 &   0.877                      &     0.901                     \\
\textit{\textbf{Comcast}}         &   1.74            &   0.712               &   0.824                      &    0.857                      \\
\textit{\textbf{Delta}}           &  2             &   0.678               &   0.776                      &      0.814                    \\
\textit{\textbf{Hulu}}            &   1.84           &   0.712                &  0.81                       &  0.844                        \\
\textit{\textbf{PlayStation}}     &  0.54             &   0.877                &   0.95                      &    0.96                     \\
\textit{\textbf{Sainsburys}}      &  1.64             &  0.743              &  0.833                      &    0.862                      \\
\textit{\textbf{Spectrum}}        &  0.42             &  0.767               &  0.857                       &   0.883                       \\
\textit{\textbf{Spotify}}         &  1.57             &   0.74                &  0.831                      &   0.862                      \\
\textit{\textbf{Sprint}}          &  0.83             &   0.825                &  0.916                       &   0.939                      \\
\textit{\textbf{Tesco}}           &   2.47            &   0.616                &   0.738                      & 0.783                        \\
\textit{\textbf{TMobile}}         &  2.1            &   0.653               &   0.765                      &    0.808                      \\
\textit{\textbf{Uber}}            &  1.68             &   0.713                &   0.813                     &     0.847                     \\
\textit{\textbf{UPS}}             &   1.04            &   0.783               &   0.897                     &   0.926                       \\
\textit{\textbf{Virgin Trains}}   &   1.67            &   0.746                &  0.819                      &   0.846                      \\
\textit{\textbf{Xbox}}            &  1.57             &   0.761                &   0.842                      &   0.87                       \\ \bottomrule
\end{tabular}
\end{table}

Table \ref{tab:no-transfer} shows the results for training the models from an initial random weight distribution, i.e., non-transfer learning. For several of the larger datasets, the classification accuracy is relatively low. It was observed that, although this is the case, the chatbot was capable of answering queries in correct English, with few obvious mistakes (examples of interaction are presented later in this section). As expected, the top-5 and top-10 accuracies are much higher, especially in situations of low loss, showing that although accuracy may be relatively low, the correct prediction often features within the top $k$ predictions. 

\begin{table*}[]
\centering
\caption{Validation loss when transfer learning (grey cells denote non-transfer).}
\label{tab:transfer-loss}
\footnotesize
\tabcolsep=0.15cm
\begin{tabular}{@{}lrrrrrrrrrrrrrrrrrrr@{}}
\toprule
                                  & \multicolumn{19}{l}{\textbf{Source}}                                                    \\ \midrule
\textbf{Target}                   & \multicolumn{1}{l}{\rotatebox{90}{\textit{\textbf{Amazon}}}} & \multicolumn{1}{l}{\rotatebox{90}{\textit{\textbf{American Air}}}} & \multicolumn{1}{l}{\rotatebox{90}{\textit{\textbf{Apple Support}}}} & \multicolumn{1}{l}{\rotatebox{90}{\textit{\textbf{British Airways}}}} & \multicolumn{1}{l}{\rotatebox{90}{\textit{\textbf{Chipotle}}}} & \multicolumn{1}{l}{\rotatebox{90}{\textit{\textbf{Comcast}}}} & \multicolumn{1}{l}{\rotatebox{90}{\textit{\textbf{Delta}}}} & \multicolumn{1}{l}{\rotatebox{90}{\textit{\textbf{Hulu}}}} & \multicolumn{1}{l}{\rotatebox{90}{\textit{\textbf{PlayStation}}}} & \multicolumn{1}{l}{\rotatebox{90}{\textit{\textbf{Sainsburys}}}} & \multicolumn{1}{l}{\rotatebox{90}{\textit{\textbf{Spectrum}}}} & \multicolumn{1}{l}{\rotatebox{90}{\textit{\textbf{Spotify}}}} & \multicolumn{1}{l}{\rotatebox{90}{\textit{\textbf{Sprint}}}} & \multicolumn{1}{l}{\rotatebox{90}{\textit{\textbf{Tesco}}}} & \multicolumn{1}{l}{\rotatebox{90}{\textit{\textbf{TMobile}}}} & \multicolumn{1}{l}{\rotatebox{90}{\textit{\textbf{Uber}}}} & \multicolumn{1}{l}{\rotatebox{90}{\textit{\textbf{UPS}}}} & \multicolumn{1}{l}{\rotatebox{90}{\textit{\textbf{Virgin Trains}}}} & \multicolumn{1}{l}{\rotatebox{90}{\textit{\textbf{Xbox}}}} \\ \cmidrule(l){2-20} 
\textit{\textbf{Amazon}}          & \cellcolor[HTML]{A5A5A5}2.3                  & 2.38                                               & 2.37                                                & 2.37                                                  & 2.36                                           & 2.36                                          & 2.38                                        & 2.36                                       & 2.35                                              & 2.37                                             & 2.36                                           & 2.36                                          & 2.35                                         & 2.38                                        & 2.37                                          & 2.36                                       & 2.35                                      & 2.36                                                & 2.36                                       \\
\textit{\textbf{\begin{tabular}[c]{@{}l@{}}American\\Air\end{tabular}}}   & 2.28                                         & \cellcolor[HTML]{A5A5A5}2.17                       & 2.19                                                & 2.13                                                  & 2.14                                           & 2.13                                          & 2.13                                        & 2.13                                       & 2.21                                              & 2.15                                             & 2.13                                           & 2.14                                          & 2.14                                         & 2.15                                        & 2.13                                          & 2.14                                       & 2.12                                      & 2.14                                                & 2.13                                       \\
\textit{\textbf{\begin{tabular}[c]{@{}l@{}}Apple\\Support\end{tabular}}}   & 2.04                                         & 1.99                                               & \cellcolor[HTML]{A5A5A5}1.93                        & 1.99                                                  & 1.98                                           & 1.98                                          & 2                                           & 1.98                                       & 1.97                                              & 1.99                                             & 1.97                                           & 1.99                                          & 1.96                                         & 2                                           & 1.98                                          & 1.99                                       & 1.97                                      & 1.99                                                & 1.97                                       \\
\textit{\textbf{\begin{tabular}[c]{@{}l@{}}British\\Airways\end{tabular}}} & 2.34                                         & 2.25                                               & 2.309                                               & \cellcolor[HTML]{A5A5A5}2.19                          & 2.25                                           & 2.24                                          & 2.24                                        & 2.24                                       & 2.23                                              & 2.26                                             & 2.24                                           & 2.26                                          & 2.24                                         & 2.23                                        & 2.25                                          & 2.26                                       & 2.22                                      & 2.25                                                & 2.26                                       \\
\textit{\textbf{Chipotle}}        & 1.15                                         & 1.1                                                & 1.16                                                & 1.09                                                  & \cellcolor[HTML]{A5A5A5}1.17                   & 1.09                                          & 1.09                                        & 1.1                                        & 1.1                                               & 1.09                                             & 1.08                                           & 1.12                                          & 1.09                                         & 1.08                                        & 1.09                                          & 1.09                                       & 1.01                                      & 1.11                                                & 1.08                                       \\
\textit{\textbf{Comcast}}         & 1.8                                          & 1.71                                               & 1.78                                                & 1.73                                                  & 1.71                                           & \cellcolor[HTML]{A5A5A5}1.74                  & 1.72                                        & 1.7                                        & 1.71                                              & 1.73                                             & 1.76                                           & 1.71                                          & 1.72                                         & 1.71                                        & 1.72                                          & 1.72                                       & 1.69                                      & 1.78                                                & 1.8                                        \\
\textit{\textbf{Delta}}           & 2.14                                         & 1.96                                               & 2.05                                                & 2.07                                                  & 2.07                                           & 2.05                                          & \cellcolor[HTML]{A5A5A5}2                   & 2.06                                       & 2.05                                              & 2.07                                             & 2.05                                           & 2.02                                          & 2.05                                         & 2.06                                        & 2.06                                          & 2.02                                       & 2.03                                      & 2.04                                                & 2.05                                       \\
\textit{\textbf{Hulu}}            & 1.88                                         & 1.84                                               & 1.91                                                & 1.81                                                  & 1.83                                           & 1.82                                          & 1.8                                         & \cellcolor[HTML]{A5A5A5}1.84               & 1.78                                              & 1.85                                             & 1.78                                           & 1.82                                          & 1.8                                          & 1.8                                         & 1.81                                          & 1.82                                       & 1.83                                      & 1.83                                                & 1.8                                        \\
\textit{\textbf{PlayStation}}     & 0.58                                         & 0.31                                               & 0.45                                                & 0.31                                                  & 0.33                                           & 0.32                                          & 0.32                                        & 0.31                                       & \cellcolor[HTML]{A5A5A5}0.54                      & 0.31                                             & 0.33                                           & 0.33                                          & 0.32                                         & 0.3                                         & 0.32                                          & 0.33                                       & 0.33                                      & 0.37                                                & 0.31                                       \\
\textit{\textbf{Sainsburys}}      & 1.76                                         & 1.62                                               & 1.72                                                & 1.62                                                  & 1.63                                           & 1.61                                          & 1.67                                        & 1.61                                       & 1.62                                              & \cellcolor[HTML]{A5A5A5}1.64                     & 1.64                                           & 1.64                                          & 1.62                                         & 1.6                                         & 1.61                                          & 1.63                                       & 1.61                                      & 1.63                                                & 1.61                                       \\
\textit{\textbf{Spectrum}}        & 1.55                                         & 1.5                                                & 1.52                                                & 1.5                                                   & 1.49                                           & 1.47                                          & 1.49                                        & 1.48                                       & 1.49                                              & 1.5                                              & \cellcolor[HTML]{A5A5A5}1.42                   & 1.49                                          & 1.47                                         & 1.46                                        & 1.45                                          & 1.49                                       & 1.46                                      & 1.41                                                & 1.49                                       \\
\textit{\textbf{Spotify}}         & 1.61                                         & 1.55                                               & 1.55                                                & 1.54                                                  & 1.55                                           & 1.55                                          & 1.55                                        & 1.54                                       & 1.52                                              & 1.54                                             & 1.52                                           & \cellcolor[HTML]{A5A5A5}1.57                  & 1.53                                         & 1.57                                        & 1.54                                          & 1.55                                       & 1.52                                      & 1.54                                                & 1.54                                       \\
\textit{\textbf{Sprint}}          & 0.98                                         & 0.6                                                & 0.85                                                & 0.6                                                   & 0.63                                           & 0.6                                           & 0.61                                        & 0.6                                        & 0.63                                              & 0.6                                              & 0.61                                           & 0.62                                          & \cellcolor[HTML]{A5A5A5}0.83                 & 0.59                                        & 0.6                                           & 0.64                                       & 0.62                                      & 0.66                                                & 0.61                                       \\
\textit{\textbf{Tesco}}           & 2.87                                         & 2.45                                               & 2.51                                                & 2.44                                                  & 2.46                                           & 2.45                                          & 2.45                                        & 2.44                                       & 2.45                                              & 2.45                                             & 2.42                                           & 2.46                                          & 2.42                                         & \cellcolor[HTML]{A5A5A5}2.47                & 2.43                                          & 2.45                                       & 2.41                                      & 2.45                                                & 2.43                                       \\
\textit{\textbf{TMobile}}         & 2.18                                         & 2.07                                               & 2.11                                                & 2.07                                                  & 2.07                                           & 2.05                                          & 2.07                                        & 2.05                                       & 2.05                                              & 2.08                                             & 2.04                                           & 2.07                                          & 2.06                                         & 2.06                                        & \cellcolor[HTML]{A5A5A5}2.1                   & 2.07                                       & 2.04                                      & 2.13                                                & 2.06                                       \\
\textit{\textbf{Uber}}            & 1.81                                         & 1.72                                               & 1.74                                                & 1.71                                                  & 1.73                                           & 1.7                                           & 1.72                                        & 1.7                                        & 1.69                                              & 1.71                                             & 1.7                                            & 1.71                                          & 1.7                                          & 1.72                                        & 1.7                                           & \cellcolor[HTML]{A5A5A5}1.68               & 1.69                                      & 1.7                                                 & 1.71                                       \\
\textit{\textbf{UPS}}             & 1.21                                         & 0.79                                               & 1.07                                                & 0.79                                                  & 0.84                                           & 0.79                                          & 0.81                                        & 0.79                                       & 0.83                                              & 0.84                                             & 0.81                                           & 0.85                                          & 0.8                                          & 0.79                                        & 0.8                                           & 0.85                                       & \cellcolor[HTML]{A5A5A5}1.04              & 0.88                                                & 0.81                                       \\
\textit{\textbf{\begin{tabular}[c]{@{}l@{}}Virgin\\Trains\end{tabular}}}   & 1.74                                         & 1.17                                               & 1.75                                                & 1.66                                                  & 1.66                                           & 1.66                                          & 1.67                                        & 1.66                                       & 1.65                                              & 1.66                                             & 1.66                                           & 1.68                                          & 1.65                                         & 1.66                                        & 1.66                                          & 1.67                                       & 1.65                                      & \cellcolor[HTML]{A5A5A5}1.67                        & 1.66                                       \\
\textit{\textbf{Xbox}}            & 1.66                                         & 1.63                                               & 1.65                                                & 1.59                                                  & 1.59                                           & 1.64                                          & 1.64                                        & 1.61                                       & 1.6                                               & 1.62                                             & 1.62                                           & 1.63                                          & 1.62                                         & 1.6                                         & 1.6                                           & 1.63                                       & 1.58                                      & 1.62                                                & \cellcolor[HTML]{A5A5A5}1.57               \\ \bottomrule
\end{tabular}
\end{table*}

The results in Table \ref{tab:transfer-loss} show the losses when transfer learning between all domains, as well as the non-transfer results highlighted in grey. It can be observed that classification losses were reduced for 13 of the 19 domains (those that did not decrease were Amazon, Apple Support, British Airways, Uber and Xbox) when transfer learning from at least one other. It is interesting to note that the largest datasets in particular were far from benefiting from transfer learning, with the learning of the majority of smaller datasets being greatly improved by transferring from other domains. This suggests that transfer of knowledge is possibly a solution to data scarcity in the customer support chatbot problem. Of the five models that lacked improvement, British Airways and Xbox were the most interesting in particular, given that transferring from other airlines (American Air, Delta) and other console manufacturers (PlayStation) did not provide improvements. Otherwise, all domains found improvement in one or more cases when learning was transferred from other domains.

\begin{table*}[]
\centering
\caption{Validation accuracy when transfer learning (grey cells denote non-transfer).}
\label{tab:transfer-acc}
\footnotesize
\tabcolsep=0.12cm
\begin{tabular}{@{}lrrrrrrrrrrrrrrrrrrr@{}}
\toprule
                                  & \multicolumn{19}{l}{\textbf{Source}}                                                    \\ \midrule
\textbf{Target}                   & \multicolumn{1}{l}{\rotatebox{90}{\textit{\textbf{Amazon}}}} & \multicolumn{1}{l}{\rotatebox{90}{\textit{\textbf{American Air}}}} & \multicolumn{1}{l}{\rotatebox{90}{\textit{\textbf{Apple Support}}}} & \multicolumn{1}{l}{\rotatebox{90}{\textit{\textbf{British Airways}}}} & \multicolumn{1}{l}{\rotatebox{90}{\textit{\textbf{Chipotle}}}} & \multicolumn{1}{l}{\rotatebox{90}{\textit{\textbf{Comcast}}}} & \multicolumn{1}{l}{\rotatebox{90}{\textit{\textbf{Delta}}}} & \multicolumn{1}{l}{\rotatebox{90}{\textit{\textbf{Hulu}}}} & \multicolumn{1}{l}{\rotatebox{90}{\textit{\textbf{PlayStation}}}} & \multicolumn{1}{l}{\rotatebox{90}{\textit{\textbf{Sainsburys}}}} & \multicolumn{1}{l}{\rotatebox{90}{\textit{\textbf{Spectrum}}}} & \multicolumn{1}{l}{\rotatebox{90}{\textit{\textbf{Spotify}}}} & \multicolumn{1}{l}{\rotatebox{90}{\textit{\textbf{Sprint}}}} & \multicolumn{1}{l}{\rotatebox{90}{\textit{\textbf{Tesco}}}} & \multicolumn{1}{l}{\rotatebox{90}{\textit{\textbf{TMobile}}}} & \multicolumn{1}{l}{\rotatebox{90}{\textit{\textbf{Uber}}}} & \multicolumn{1}{l}{\rotatebox{90}{\textit{\textbf{UPS}}}} & \multicolumn{1}{l}{\rotatebox{90}{\textit{\textbf{Virgin Trains}}}} & \multicolumn{1}{l}{\rotatebox{90}{\textit{\textbf{Xbox}}}} \\ \cmidrule(l){2-20} 
\textit{\textbf{Amazon}}                                                    & \cellcolor[HTML]{A5A5A5}0.602                & 0.6                                                & 0.595                                               & 0.6                                                   & 0.6                                            & 0.6                                           & 0.6                                         & 0.6                                        & 0.601                                             & 0.6                                              & 0.6                                            & 0.6                                           & 0.601                                        & 0.6                                         & 0.599                                         & 0.598                                      & 0.601                                     & 0.598                                               & 0.6                                        \\
\textit{\textbf{\begin{tabular}[c]{@{}l@{}}American \\ Air\end{tabular}}}   & 0.644                                        & \cellcolor[HTML]{A5A5A5}0.645                      & 0.63                                                & 0.647                                                 & 0.643                                          & 0.646                                         & 0.647                                       & 0.645                                      & 0.652                                             & 0.646                                            & 0.644                                          & 0.644                                         & 0.648                                        & 0.647                                       & 0.646                                         & 0.644                                      & 0.648                                     & 0.643                                               & 0.652                                      \\
\textit{\textbf{\begin{tabular}[c]{@{}l@{}}Apple\\Support\end{tabular}}}  & 0.632                                        & 0.639                                              & \cellcolor[HTML]{A5A5A5}0.642                       & 0.64                                                  & 0.638                                          & 0.639                                         & 0.64                                        & 0.64                                       & 0.641                                             & 0.639                                            & 0.64                                           & 0.638                                         & 0.641                                        & 0.639                                       & 0.64                                          & 0.638                                      & 0.64                                      & 0.637                                               & 0.639                                      \\
\textit{\textbf{\begin{tabular}[c]{@{}l@{}}British\\Airways\end{tabular}}} & 0.648                                        & 0.663                                              & 0.634                                               & \cellcolor[HTML]{A5A5A5}0.658                         & 0.658                                          & 0.663                                         & 0.663                                       & 0.665                                      & 0.66                                              & 0.664                                            & 0.66                                           & 0.659                                         & 0.662                                        & 0.665                                       & 0.663                                         & 0.657                                      & 0.662                                     & 0.657                                               & 0.663                                      \\
\textit{\textbf{Chipotle}}                                                  & 0.81                                         & 0.842                                              & 0.815                                               & 0.842                                                 & \cellcolor[HTML]{A5A5A5}0.808                  & 0.842                                         & 0.842                                       & 0.843                                      & 0.839                                             & 0.841                                            & 0.839                                          & 0.839                                         & 0.838                                        & 0.844                                       & 0.842                                         & 0.838                                      & 0.839                                     & 0.835                                               & 0.839                                      \\
\textit{\textbf{Comcast}}                                                   & 0.7                                          & 0.714                                              & 0.693                                               & 0.713                                                 & 0.711                                          & \cellcolor[HTML]{A5A5A5}0.712                 & 0.712                                       & 0.714                                      & 0.708                                             & 0.712                                            & 0.727                                          & 0.71                                          & 0.727                                        & 0.716                                       & 0.714                                         & 0.708                                      & 0.708                                     & 0.719                                               & 0.72                                       \\
\textit{\textbf{Delta}}                                                     & 0.68                                         & 0.68                                               & 0.668                                               & 0.687                                                 & 0.683                                          & 0.687                                         & \cellcolor[HTML]{A5A5A5}0.678               & 0.686                                      & 0.684                                             & 0.687                                            & 0.686                                          & 0.681                                         & 0.685                                        & 0.687                                       & 0.687                                         & 0.68                                       & 0.686                                     & 0.682                                               & 0.686                                      \\
\textit{\textbf{Hulu}}                                                      & 0.69                                         & 0.7                                                & 0.685                                               & 0.713                                                 & 0.71                                           & 0.714                                         & 0.712                                       & \cellcolor[HTML]{A5A5A5}0.712              & 0.714                                             & 0.714                                            & 0.713                                          & 0.71                                          & 0.712                                        & 0.714                                       & 0.713                                         & 0.701                                      & 0.71                                      & 0.704                                               & 0.714                                      \\
\textit{\textbf{PlayStation}}                                               & 0.865                                        & 0.928                                              & 0.892                                               & 0.927                                                 & 0.923                                          & 0.925                                         & 0.924                                       & 0.926                                      & \cellcolor[HTML]{A5A5A5}0.877                     & 0.926                                            & 0.923                                          & 0.924                                         & 0.924                                        & 0.929                                       & 0.927                                         & 0.921                                      & 0.923                                     & 0.916                                               & 0.93                                       \\
\textit{\textbf{Sainsburys}}                                                & 0.731                                        & 0.759                                              & 0.733                                               & 0.757                                                 & 0.764                                          & 0.757                                         & 0.763                                       & 0.769                                      & 0.766                                             & \cellcolor[HTML]{A5A5A5}0.743                    & 0.762                                          & 0.751                                         & 0.767                                        & 0.761                                       & 0.759                                         & 0.752                                      & 0.756                                     & 0.762                                               & 0.768                                      \\
\textit{\textbf{Spectrum}}                                                  & 0.743                                        & 0.766                                              & 0.748                                               & 0.765                                                 & 0.761                                          & 0.768                                         & 0.766                                       & 0.768                                      & 0.762                                             & 0.767                                            & \cellcolor[HTML]{A5A5A5}0.767                  & 0.762                                         & 0.765                                        & 0.768                                       & 0.765                                         & 0.762                                      & 0.762                                     & 0.774                                               & 0.766                                      \\
\textit{\textbf{Spotify}}                                                   & 0.737                                        & 0.742                                              & 0.732                                               & 0.742                                                 & 0.739                                          & 0.741                                         & 0.74                                        & 0.742                                      & 0.741                                             & 0.741                                            & 0.742                                          & \cellcolor[HTML]{A5A5A5}0.74                  & 0.742                                        & 0.74                                        & 0.742                                         & 0.74                                       & 0.741                                     & 0.739                                               & 0.742                                      \\
\textit{\textbf{Sprint}}                                                    & 0.791                                        & 0.882                                              & 0.827                                               & 0.883                                                 & 0.877                                          & 0.882                                         & 0.881                                       & 0.882                                      & 0.876                                             & 0.882                                            & 0.879                                          & 0.877                                         & \cellcolor[HTML]{A5A5A5}0.825                & 0.883                                       & 0.881                                         & 0.871                                      & 0.88                                      & 0.871                                               & 0.882                                      \\
\textit{\textbf{Tesco}}                                                     & 0.613                                        & 0.615                                              & 0.596                                               & 0.616                                                 & 0.612                                          & 0.615                                         & 0.615                                       & 0.615                                      & 0.616                                             & 0.616                                            & 0.614                                          & 0.614                                         & 0.615                                        & \cellcolor[HTML]{A5A5A5}0.616               & 0.616                                         & 0.613                                      & 0.615                                     & 0.61                                                & 0.616                                      \\
\textit{\textbf{TMobile}}                                                   & 0.65                                         & 0.664                                              & 0.637                                               & 0.656                                                 & 0.653                                          & 0.657                                         & 0.655                                       & 0.657                                      & 0.653                                             & 0.656                                            & 0.657                                          & 0.657                                         & 0.662                                        & 0.658                                       & \cellcolor[HTML]{A5A5A5}0.653                 & 0.653                                      & 0.655                                     & 0.66                                                & 0.656                                      \\
\textit{\textbf{Uber}}                                                      & 0.712                                        & 0.72                                               & 0.702                                               & 0.716                                                 & 0.713                                          & 0.714                                         & 0.713                                       & 0.715                                      & 0.712                                             & 0.714                                            & 0.714                                          & 0.714                                         & 0.713                                        & 0.715                                       & 0.716                                         & \cellcolor[HTML]{A5A5A5}0.713              & 0.714                                     & 0.711                                               & 0.714                                      \\
\textit{\textbf{UPS}}                                                       & 0.741                                        & 0.838                                              & 0.769                                               & 0.843                                                 & 0.832                                          & 0.84                                          & 0.839                                       & 0.84                                       & 0.835                                             & 0.839                                            & 0.835                                          & 0.829                                         & 0.84                                         & 0.841                                       & 0.841                                         & 0.826                                      & \cellcolor[HTML]{A5A5A5}0.783             & 0.825                                               & 0.841                                      \\
\textit{\textbf{\begin{tabular}[c]{@{}l@{}}Virgin \\ Trains\end{tabular}}}  & 0.751                                        & 0.744                                              & 0.722                                               & 0.744                                                 & 0.739                                          & 0.739                                         & 0.742                                       & 0.743                                      & 0.738                                             & 0.741                                            & 0.74                                           & 0.739                                         & 0.74                                         & 0.744                                       & 0.742                                         & 0.738                                      & 0.738                                     & \cellcolor[HTML]{A5A5A5}0.746                       & 0.742                                      \\
\textit{\textbf{Xbox}}                                                      & 0.735                                        & 0.737                                              & 0.732                                               & 0.751                                                 & 0.758                                          & 0.737                                         & 0.737                                       & 0.771                                      & 0.767                                             & 0.772                                            & 0.769                                          & 0.734                                         & 0.771                                        & 0.772                                       & 0.737                                         & 0.766                                      & 0.77                                      & 0.763                                               & \cellcolor[HTML]{A5A5A5}0.761              \\ \bottomrule
\end{tabular}
\end{table*}

In terms of classification accuracy, Table \ref{tab:transfer-acc} shows the accuracy of predictions regarding only the top-1 prediction. Of the 19 domains, 15 experienced higher classification accuracy when transferring knowledge from at least one other domain. Several instances were slight increases, there were experiments that showed a much larger increase in ability when transfer learning. Three main examples of this can be seen, transfer of knowledge from Tesco to Chipotle leads accuracy to rise from 0.808 to 0.844, secondly, transferring from Tesco to Sprint leads the accuracy to rise from 0.825 to 0.883. The most interesting example, though, is when a transfer of knowledge is performed between two similar domains, Xbox to PlayStation, which causes the accuracy to increase from 0.877 to 0.93. This is particularly interesting since the problems experienced by users of these two services tend to be similar, albeit on different platforms. The UPS chatbot was improved by transferring from every other domain except for Amazon; Tesco, TMobile, and Xbox transfer learning caused the accuracy to rise from 0.783 to 0.841. A similar observation can be made from the top-$k$ results, with transfer learning aiding in correctly predicting the next object in the sequence to be contained within the top 5 and 10 predictions. 

\begin{table*}[]
\centering
\caption{Validation top-5 accuracy when transfer learning (grey cells denote non-transfer).}
\label{tab:transfer-top-5}
\footnotesize
\tabcolsep=0.12cm
\begin{tabular}{@{}lrrrrrrrrrrrrrrrrrrr@{}}
\toprule
                                  & \multicolumn{19}{l}{\textbf{Source}}                                                    \\ \midrule
\textbf{Target}                   & \multicolumn{1}{l}{\rotatebox{90}{\textit{\textbf{Amazon}}}} & \multicolumn{1}{l}{\rotatebox{90}{\textit{\textbf{American Air}}}} & \multicolumn{1}{l}{\rotatebox{90}{\textit{\textbf{Apple Support}}}} & \multicolumn{1}{l}{\rotatebox{90}{\textit{\textbf{British Airways}}}} & \multicolumn{1}{l}{\rotatebox{90}{\textit{\textbf{Chipotle}}}} & \multicolumn{1}{l}{\rotatebox{90}{\textit{\textbf{Comcast}}}} & \multicolumn{1}{l}{\rotatebox{90}{\textit{\textbf{Delta}}}} & \multicolumn{1}{l}{\rotatebox{90}{\textit{\textbf{Hulu}}}} & \multicolumn{1}{l}{\rotatebox{90}{\textit{\textbf{PlayStation}}}} & \multicolumn{1}{l}{\rotatebox{90}{\textit{\textbf{Sainsburys}}}} & \multicolumn{1}{l}{\rotatebox{90}{\textit{\textbf{Spectrum}}}} & \multicolumn{1}{l}{\rotatebox{90}{\textit{\textbf{Spotify}}}} & \multicolumn{1}{l}{\rotatebox{90}{\textit{\textbf{Sprint}}}} & \multicolumn{1}{l}{\rotatebox{90}{\textit{\textbf{Tesco}}}} & \multicolumn{1}{l}{\rotatebox{90}{\textit{\textbf{TMobile}}}} & \multicolumn{1}{l}{\rotatebox{90}{\textit{\textbf{Uber}}}} & \multicolumn{1}{l}{\rotatebox{90}{\textit{\textbf{UPS}}}} & \multicolumn{1}{l}{\rotatebox{90}{\textit{\textbf{Virgin Trains}}}} & \multicolumn{1}{l}{\rotatebox{90}{\textit{\textbf{Xbox}}}} \\ \cmidrule(l){2-20} 
\textit{\textbf{Amazon}}                                                    & \cellcolor[HTML]{A5A5A5}0.73                 & 0.725                                              & 0.724                                               & 0.724                                                 & 0.724                                          & 0.725                                         & 0.725                                       & 0.725                                      & 0.727                                             & 0.725                                            & 0.726                                          & 0.726                                         & 0.726                                        & 0.724                                       & 0.725                                         & 0.724                                      & 0.727                                     & 0.725                                               & 0.726                                      \\
\textit{\textbf{\begin{tabular}[c]{@{}l@{}}American \\ Air\end{tabular}}}   & 0.755                                        & \cellcolor[HTML]{A5A5A5}0.754                      & 0.749                                               & 0.756                                                 & 0.754                                          & 0.755                                         & 0.757                                       & 0.755                                      & 0.756                                             & 0.754                                            & 0.755                                          & 0.756                                         & 0.758                                        & 0.755                                       & 0.756                                         & 0.755                                      & 0.757                                     & 0.753                                               & 0.755                                      \\
\textit{\textbf{\begin{tabular}[c]{@{}l@{}}Apple \\ Support\end{tabular}}}  & 0.77                                         & 0.773                                              & \cellcolor[HTML]{A5A5A5}0.778                       & 0.773                                                 & 0.773                                          & 0.773                                         & 0.772                                       & 0.773                                      & 0.775                                             & 0.773                                            & 0.774                                          & 0.774                                         & 0.775                                        & 0.772                                       & 0.774                                         & 0.772                                      & 0.774                                     & 0.773                                               & 0.774                                      \\
\textit{\textbf{\begin{tabular}[c]{@{}l@{}}British\\ Airways\end{tabular}}} & 0.759                                        & 0.766                                              & 0.75                                                & \cellcolor[HTML]{A5A5A5}0.764                         & 0.759                                          & 0.763                                         & 0.763                                       & 0.764                                      & 0.76                                              & 0.762                                            & 0.762                                          & 0.761                                         & 0.762                                        & 0.764                                       & 0.762                                         & 0.761                                      & 0.762                                     & 0.759                                               & 0.77                                       \\
\textit{\textbf{Chipotle}}                                                  & 0.89                                         & 0.897                                              & 0.884                                               & 0.9                                                   & \cellcolor[HTML]{A5A5A5}0.877                  & 0.896                                         & 0.896                                       & 0.895                                      & 0.893                                             & 0.895                                            & 0.893                                          & 0.895                                         & 0.895                                        & 0.897                                       & 0.896                                         & 0.895                                      & 0.894                                     & 0.892                                               & 0.9                                        \\
\textit{\textbf{Comcast}}                                                   & 0.824                                        & 0.826                                              & 0.82                                                & 0.824                                                 & 0.822                                          & \cellcolor[HTML]{A5A5A5}0.824                 & 0.825                                       & 0.825                                      & 0.823                                             & 0.824                                            & 0.829                                          & 0.822                                         & 0.832                                        & 0.826                                       & 0.826                                         & 0.823                                      & 0.825                                     & 0.82                                                & 0.828                                      \\
\textit{\textbf{Delta}}                                                     & 0.776                                        & 0.78                                               & 0.769                                               & 0.778                                                 & 0.776                                          & 0.778                                         & \cellcolor[HTML]{A5A5A5}0.776               & 0.778                                      & 0.777                                             & 0.778                                            & 0.778                                          & 0.776                                         & 0.776                                        & 0.778                                       & 0.779                                         & 0.775                                      & 0.779                                     & 0.776                                               & 0.778                                      \\
\textit{\textbf{Hulu}}                                                      & 0.81                                         & 0.81                                               & 0.799                                               & 0.812                                                 & 0.809                                          & 0.811                                         & 0.811                                       & \cellcolor[HTML]{A5A5A5}0.81               & 0.811                                             & 0.81                                             & 0.811                                          & 0.81                                          & 0.812                                        & 0.812                                       & 0.812                                         & 0.804                                      & 0.81                                      & 0.806                                               & 0.81                                       \\
\textit{\textbf{PlayStation}}                                               & 0.94                                         & 0.97                                               & 0.956                                               & 0.97                                                  & 0.969                                          & 0.97                                          & 0.97                                        & 0.97                                       & \cellcolor[HTML]{A5A5A5}0.95                      & 0.97                                             & 0.968                                          & 0.969                                         & 0.97                                         & 0.972                                       & 0.97                                          & 0.968                                      & 0.969                                     & 0.965                                               & 0.97                                       \\
\textit{\textbf{Sainsburys}}                                                & 0.828                                        & 0.84                                               & 0.829                                               & 0.838                                                 & 0.842                                          & 0.837                                         & 0.838                                       & 0.842                                      & 0.841                                             & \cellcolor[HTML]{A5A5A5}0.833                    & 0.84                                           & 0.836                                         & 0.843                                        & 0.84                                        & 0.838                                         & 0.836                                      & 0.838                                     & 0.839                                               & 0.843                                      \\
\textit{\textbf{Spectrum}}                                                  & 0.848                                        & 0.856                                              & 0.884                                               & 0.855                                                 & 0.852                                          & 0.855                                         & 0.855                                       & 0.856                                      & 0.854                                             & 0.854                                            & \cellcolor[HTML]{A5A5A5}0.857                  & 0.854                                         & 0.854                                        & 0.856                                       & 0.855                                         & 0.854                                      & 0.853                                     & 0.859                                               & 0.855                                      \\
\textit{\textbf{Spotify}}                                                   & 0.832                                        & 0.833                                              & 0.83                                                & 0.832                                                 & 0.83                                           & 0.831                                         & 0.831                                       & 0.831                                      & 0.831                                             & 0.831                                            & 0.833                                          & \cellcolor[HTML]{A5A5A5}0.831                 & 0.832                                        & 0.831                                       & 0.831                                         & 0.831                                      & 0.832                                     & 0.83                                                & 0.833                                      \\
\textit{\textbf{Sprint}}                                                    & 0.893                                        & 0.942                                              & 0.913                                               & 0.941                                                 & 0.939                                          & 0.94                                          & 0.941                                       & 0.942                                      & 0.937                                             & 0.942                                            & 0.939                                          & 0.938                                         & \cellcolor[HTML]{A5A5A5}0.916                & 0.943                                       & 0.942                                         & 0.936                                      & 0.94                                      & 0.934                                               & 0.941                                      \\
\textit{\textbf{Tesco}}                                                     & 0.735                                        & 0.735                                              & 0.725                                               & 0.734                                                 & 0.731                                          & 0.734                                         & 0.733                                       & 0.734                                      & 0.735                                             & 0.733                                            & 0.734                                          & 0.734                                         & 0.734                                        & \cellcolor[HTML]{A5A5A5}0.738               & 0.733                                         & 0.732                                      & 0.735                                     & 0.731                                               & 0.735                                      \\
\textit{\textbf{TMobile}}                                                   & 0.77                                         & 0.772                                              & 0.759                                               & 0.767                                                 & 0.765                                          & 0.768                                         & 0.768                                       & 0.768                                      & 0.767                                             & 0.767                                            & 0.768                                          & 0.767                                         & 0.772                                        & 0.768                                       & \cellcolor[HTML]{A5A5A5}0.765                 & 0.767                                      & 0.768                                     & 0.766                                               & 0.767                                      \\
\textit{\textbf{Uber}}                                                      & 0.812                                        & 0.813                                              & 0.81                                                & 0.814                                                 & 0.812                                          & 0.832                                         & 0.813                                       & 0.813                                      & 0.813                                             & 0.813                                            & 0.813                                          & 0.813                                         & 0.813                                        & 0.813                                       & 0.814                                         & \cellcolor[HTML]{A5A5A5}0.813              & 0.813                                     & 0.811                                               & 0.814                                      \\
\textit{\textbf{UPS}}                                                       & 0.871                                        & 0.923                                              & 0.888                                               & 0.926                                                 & 0.92                                           & 0.925                                         & 0.924                                       & 0.925                                      & 0.912                                             & 0.923                                            & 0.922                                          & 0.918                                         & 0.924                                        & 0.925                                       & 0.925                                         & 0.918                                      & \cellcolor[HTML]{A5A5A5}0.897             & 0.915                                               & 0.923                                      \\
\textit{\textbf{\begin{tabular}[c]{@{}l@{}}Virgin \\ Trains\end{tabular}}}  & 0.82                                         & 0.817                                              & 0.805                                               & 0.817                                                 & 0.814                                          & 0.814                                         & 0.815                                       & 0.816                                      & 0.813                                             & 0.815                                            & 0.815                                          & 0.814                                         & 0.815                                        & 0.817                                       & 0.816                                         & 0.814                                      & 0.814                                     & \cellcolor[HTML]{A5A5A5}0.819                       & 0.816                                      \\
\textit{\textbf{Xbox}}                                                      & 0.83                                         & 0.827                                              & 0.83                                                & 0.835                                                 & 0.838                                          & 0.826                                         & 0.826                                       & 0.844                                      & 0.845                                             & 0.845                                            & 0.843                                          & 0.825                                         & 0.843                                        & 0.845                                       & 0.828                                         & 0.843                                      & 0.843                                     & 0.842                                               & \cellcolor[HTML]{A5A5A5}0.842              \\ \bottomrule
\end{tabular}
\end{table*}

\begin{table*}[]
\centering
\caption{Validation top-10 accuracy when transfer learning (grey cells denote non-transfer).}
\label{tab:transfer-top-10}
\footnotesize
\tabcolsep=0.12cm
\begin{tabular}{@{}lrrrrrrrrrrrrrrrrrrr@{}}
\toprule
                                  & \multicolumn{19}{l}{\textbf{Source}}                                                    \\ \midrule
\textbf{Target}                   & \multicolumn{1}{l}{\rotatebox{90}{\textit{\textbf{Amazon}}}} & \multicolumn{1}{l}{\rotatebox{90}{\textit{\textbf{American Air}}}} & \multicolumn{1}{l}{\rotatebox{90}{\textit{\textbf{Apple Support}}}} & \multicolumn{1}{l}{\rotatebox{90}{\textit{\textbf{British Airways}}}} & \multicolumn{1}{l}{\rotatebox{90}{\textit{\textbf{Chipotle}}}} & \multicolumn{1}{l}{\rotatebox{90}{\textit{\textbf{Comcast}}}} & \multicolumn{1}{l}{\rotatebox{90}{\textit{\textbf{Delta}}}} & \multicolumn{1}{l}{\rotatebox{90}{\textit{\textbf{Hulu}}}} & \multicolumn{1}{l}{\rotatebox{90}{\textit{\textbf{PlayStation}}}} & \multicolumn{1}{l}{\rotatebox{90}{\textit{\textbf{Sainsburys}}}} & \multicolumn{1}{l}{\rotatebox{90}{\textit{\textbf{Spectrum}}}} & \multicolumn{1}{l}{\rotatebox{90}{\textit{\textbf{Spotify}}}} & \multicolumn{1}{l}{\rotatebox{90}{\textit{\textbf{Sprint}}}} & \multicolumn{1}{l}{\rotatebox{90}{\textit{\textbf{Tesco}}}} & \multicolumn{1}{l}{\rotatebox{90}{\textit{\textbf{TMobile}}}} & \multicolumn{1}{l}{\rotatebox{90}{\textit{\textbf{Uber}}}} & \multicolumn{1}{l}{\rotatebox{90}{\textit{\textbf{UPS}}}} & \multicolumn{1}{l}{\rotatebox{90}{\textit{\textbf{Virgin Trains}}}} & \multicolumn{1}{l}{\rotatebox{90}{\textit{\textbf{Xbox}}}} \\ \cmidrule(l){2-20} 
\textit{\textbf{Amazon}}                                                    & \cellcolor[HTML]{A5A5A5}0.779                & 0.774                                              & 0.774                                               & 0.774                                                 & 0.773                                          & 0.775                                         & 0.775                                       & 0.775                                      & 0.776                                             & 0.774                                            & 0.775                                          & 0.775                                         & 0.775                                        & 0.773                                       & 0.774                                         & 0.774                                      & 0.776                                     & 0.774                                               & 0.775                                      \\
\textit{\textbf{\begin{tabular}[c]{@{}l@{}}American \\ Air\end{tabular}}}   & 0.8                                          & \cellcolor[HTML]{A5A5A5}0.8                        & 0.796                                               & 0.8                                                   & 0.798                                          & 0.799                                         & 0.8                                         & 0.799                                      & 0.798                                             & 0.796                                            & 0.799                                          & 0.799                                         & 0.801                                        & 0.797                                       & 0.799                                         & 0.798                                      & 0.802                                     & 0.797                                               & 0.797                                      \\
\textit{\textbf{\begin{tabular}[c]{@{}l@{}}Apple \\ Support\end{tabular}}}  & 0.82                                         & 0.819                                              & \cellcolor[HTML]{A5A5A5}0.826                       & 0.82                                                  & 0.82                                           & 0.819                                         & 0.819                                       & 0.82                                       & 0.821                                             & 0.82                                             & 0.82                                           & 0.821                                         & 0.822                                        & 0.819                                       & 0.821                                         & 0.819                                      & 0.821                                     & 0.82                                                & 0.82                                       \\
\textit{\textbf{\begin{tabular}[c]{@{}l@{}}British\\ Airways\end{tabular}}} & 0.81                                         & 0.804                                              & 0.795                                               & \cellcolor[HTML]{A5A5A5}0.802                         & 0.797                                          & 0.801                                         & 0.802                                       & 0.802                                      & 0.799                                             & 0.8                                              & 0.801                                          & 0.801                                         & 0.8                                          & 0.802                                       & 0.801                                         & 0.8                                        & 0.8                                       & 0.799                                               & 0.82                                       \\
\textit{\textbf{Chipotle}}                                                  & 0.91                                         & 0.915                                              & 0.909                                               & 0.915                                                 & \cellcolor[HTML]{A5A5A5}0.901                  & 0.915                                         & 0.914                                       & 0.913                                      & 0.912                                             & 0.914                                            & 0.912                                          & 0.914                                         & 0.913                                        & 0.915                                       & 0.914                                         & 0.914                                      & 0.912                                     & 0.911                                               & 0.914                                      \\
\textit{\textbf{Comcast}}                                                   & 0.86                                         & 0.859                                              & 0.858                                               & 0.858                                                 & 0.857                                          & \cellcolor[HTML]{A5A5A5}0.857                 & 0.859                                       & 0.859                                      & 0.857                                             & 0.858                                            & 0.858                                          & 0.857                                         & 0.862                                        & 0.858                                       & 0.86                                          & 0.858                                      & 0.859                                     & 0.857                                               & 0.858                                      \\
\textit{\textbf{Delta}}                                                     & 0.814                                        & 0.817                                              & 0.81                                                & 0.814                                                 & 0.813                                          & 0.815                                         & \cellcolor[HTML]{A5A5A5}0.814               & 0.814                                      & 0.814                                             & 0.814                                            & 0.813                                          & 0.814                                         & 0.813                                        & 0.815                                       & 0.815                                         & 0.813                                      & 0.816                                     & 0.813                                               & 0.814                                      \\
\textit{\textbf{Hulu}}                                                      & 0.842                                        & 0.84                                               & 0.841                                               & 0.847                                                 & 0.846                                          & 0.846                                         & 0.847                                       & \cellcolor[HTML]{A5A5A5}0.844              & 0.846                                             & 0.846                                            & 0.847                                          & 0.846                                         & 0.848                                        & 0.85                                        & 0.848                                         & 0.844                                      & 0.846                                     & 0.843                                               & 0.845                                      \\
\textit{\textbf{PlayStation}}                                               & 0.961                                        & 0.981                                              & 0.973                                               & 0.98                                                  & 0.98                                           & 0.981                                         & 0.98                                        & 0.981                                      & \cellcolor[HTML]{A5A5A5}0.96                      & 0.98                                             & 0.979                                          & 0.98                                          & 0.98                                         & 0.982                                       & 0.981                                         & 0.98                                       & 0.98                                      & 0.977                                               & 0.981                                      \\
\textit{\textbf{Sainsburys}}                                                & 0.861                                        & 0.866                                              & 0.861                                               & 0.865                                                 & 0.868                                          & 0.864                                         & 0.866                                       & 0.868                                      & 0.866                                             & \cellcolor[HTML]{A5A5A5}0.862                    & 0.866                                          & 0.864                                         & 0.868                                        & 0.867                                       & 0.865                                         & 0.863                                      & 0.864                                     & 0.866                                               & 0.868                                      \\
\textit{\textbf{Spectrum}}                                                  & 0.881                                        & 0.884                                              & 0.883                                               & 0.882                                                 & 0.879                                          & 0.882                                         & 0.882                                       & 0.883                                      & 0.881                                             & 0.88                                             & \cellcolor[HTML]{A5A5A5}0.883                  & 0.881                                         & 0.881                                        & 0.883                                       & 0.883                                         & 0.882                                      & 0.881                                     & 0.885                                               & 0.882                                      \\
\textit{\textbf{Spotify}}                                                   & 0.865                                        & 0.864                                              & 0.864                                               & 0.853                                                 & 0.862                                          & 0.862                                         & 0.862                                       & 0.863                                      & 0.863                                             & 0.862                                            & 0.864                                          & \cellcolor[HTML]{A5A5A5}0.862                 & 0.863                                        & 0.861                                       & 0.863                                         & 0.863                                      & 0.863                                     & 0.862                                               & 0.863                                      \\
\textit{\textbf{Sprint}}                                                    & 0.924                                        & 0.958                                              & 0.94                                                & 0.957                                                 & 0.955                                          & 0.956                                         & 0.957                                       & 0.957                                      & 0.954                                             & 0.957                                            & 0.955                                          & 0.955                                         & \cellcolor[HTML]{A5A5A5}0.939                & 0.958                                       & 0.958                                         & 0.954                                      & 0.955                                     & 0.952                                               & 0.957                                      \\
\textit{\textbf{Tesco}}                                                     & 0.782                                        & 0.8                                                & 0.776                                               & 0.778                                                 & 0.776                                          & 0.778                                         & 0.778                                       & 0.778                                      & 0.781                                             & 0.778                                            & 0.779                                          & 0.78                                          & 0.779                                        & \cellcolor[HTML]{A5A5A5}0.783               & 0.778                                         & 0.778                                      & 0.78                                      & 0.777                                               & 0.78                                       \\
\textit{\textbf{TMobile}}                                                   & 0.811                                        & 0.813                                              & 0.806                                               & 0.811                                                 & 0.808                                          & 0.811                                         & 0.811                                       & 0.811                                      & 0.81                                              & 0.809                                            & 0.811                                          & 0.811                                         & 0.814                                        & 0.811                                       & \cellcolor[HTML]{A5A5A5}0.808                 & 0.81                                       & 0.811                                     & 0.808                                               & 0.809                                      \\
\textit{\textbf{Uber}}                                                      & 0.85                                         & 0.847                                              & 0.845                                               & 0.847                                                 & 0.846                                          & 0.847                                         & 0.846                                       & 0.846                                      & 0.847                                             & 0.846                                            & 0.847                                          & 0.846                                         & 0.846                                        & 0.847                                       & 0.847                                         & \cellcolor[HTML]{A5A5A5}0.847              & 0.847                                     & 0.846                                               & 0.847                                      \\
\textit{\textbf{UPS}}                                                       & 0.91                                         & 0.945                                              & 0.923                                               & 0.947                                                 & 0.943                                          & 0.946                                         & 0.945                                       & 0.946                                      & 0.943                                             & 0.945                                            & 0.944                                          & 0.941                                         & 0.946                                        & 0.946                                       & 0.946                                         & 0.942                                      & \cellcolor[HTML]{A5A5A5}0.926             & 0.939                                               & 0.945                                      \\
\textit{\textbf{\begin{tabular}[c]{@{}l@{}}Virgin \\ Trains\end{tabular}}}  & 0.85                                         & 0.847                                              & 0.84                                                & 0.847                                                 & 0.845                                          & 0.845                                         & 0.846                                       & 0.846                                      & 0.844                                             & 0.845                                            & 0.845                                          & 0.844                                         & 0.845                                        & 0.847                                       & 0.847                                         & 0.845                                      & 0.845                                     & \cellcolor[HTML]{A5A5A5}0.846                       & 0.846                                      \\
\textit{\textbf{Xbox}}                                                      & 0.87                                         & 0.861                                              & 0.865                                               & 0.867                                                 & 0.866                                          & 0.858                                         & 0.86                                        & 0.87                                       & 0.871                                             & 0.871                                            & 0.87                                           & 0.859                                         & 0.87                                         & 0.87                                        & 0.86                                          & 0.87                                       & 0.869                                     & 0.869                                               & \cellcolor[HTML]{A5A5A5}0.87               \\ \bottomrule
\end{tabular}
\end{table*}

Although losses were not reduced, and so errors were considered more severe, the accuracy was improved for models such as the Xbox Support chatbot when transfer learning. For example, rising from 0.761 to 0.772 when transferring weights from the Sainsburys and Tesco models. It is worth noting that these two biggest improvements came from the transfer of knowledge from two large British retailers. The best model for the PlayStation chatbot in terms of accuracy is observed to be that which is transferred from Xbox (0.877 to 0.93) which is particularly interesting due to their similar domains. This was also the case for the top-5 and top-10 accuracy observations, although other models also achieved this score (0.96 to 0.981). Results such as this suggest that in addition to transferring knowledge of the English language and conversations in general, that there is also the possibility that specific domain knowledge can be carried over as well, in some cases. To expand on the accuracy metrics, an exploration of the top-$k$ predictions is performed. Table \ref{tab:transfer-top-5} shows the accuracy when $k=5$, and \ref{tab:transfer-top-10} shows the accuracy when $k=10$. A similar pattern is observed to the prior experiments, the majority of domains experience an increase in metrics when the transfer of knowledge is performed from at least one other domain. This is the case for 16/19 domains with regard to $k=5$ and 17/19 for $k=10$, the two domains that did not experience increases stay the same between the two, while transfer learning to Tesco when $k=10$ is improved by transfer learning from American Air (0.783 to 0.8). A particularly large increase that stands out from many of the experiments is when learning is transferred from the Apple Support chatbot to the Spectrum chatbot, with $k=5$ accuracy rising from 0.857 to 0.884.

\begin{figure}
    \centering
    \includegraphics{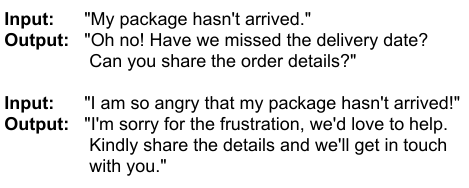}
    \caption{Example of two interactions with the chatbot trained on Amazon data. Note that although the problem is the same, the response changes due to emotion.}
    \label{fig:amazon-convo}
\end{figure}
\begin{figure}
    \centering
    \includegraphics{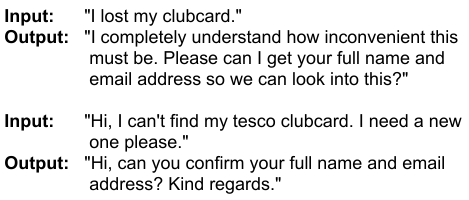}
    \caption{Example of two similar interactions with the Tesco chatbot.}
    \label{fig:tesco-convo}
\end{figure}

Examples of communication with the chatbot can be seen in Figures \ref{fig:amazon-convo} (Amazon) and \ref{fig:tesco-convo} (Tesco). Punctuation has been inferred manually for readability. Note that, although the same query has been asked, the response is tailored to the input. This can be seen especially in the Amazon examples, where a friendly tone in the first example soon changes to being apologetic and understanding once the input is edited to express negative emotions towards the situation. The Tesco example shows a different problem, since losing a membership card is a concrete problem whereas an order not arriving yet may be a non-issue. An interesting nuance can be found in the responses if they are compared, in the first instance, ``I lost my clubcard." is relatively emotionless and states the issue, with the chatbot responding that support understands how convenient the issue must be. In the second, more polite input, ``Hi, I can't find my clubcard. I need a new one please." is responded to in kind, with the chatbot choosing to wrap the solution between  ``Hi," and  ``Kind regards.". These examples of nuance show that a learning-based approach from real conversations can introduce a more natural feel to the interaction, as opposed to static responses with solutions. This achievement was one of the two major goals of this study, effectively to explore how approaches can move away from the expert system-like solution-responses and more towards a natural conversation with a machine as would be performed with another human being.

\subsection{Consumer Robot Feasibility and Implementation}
\label{sec:robots}
\label{subsec:implementation}
\begin{figure}
    \centering
    \includegraphics[scale=0.35]{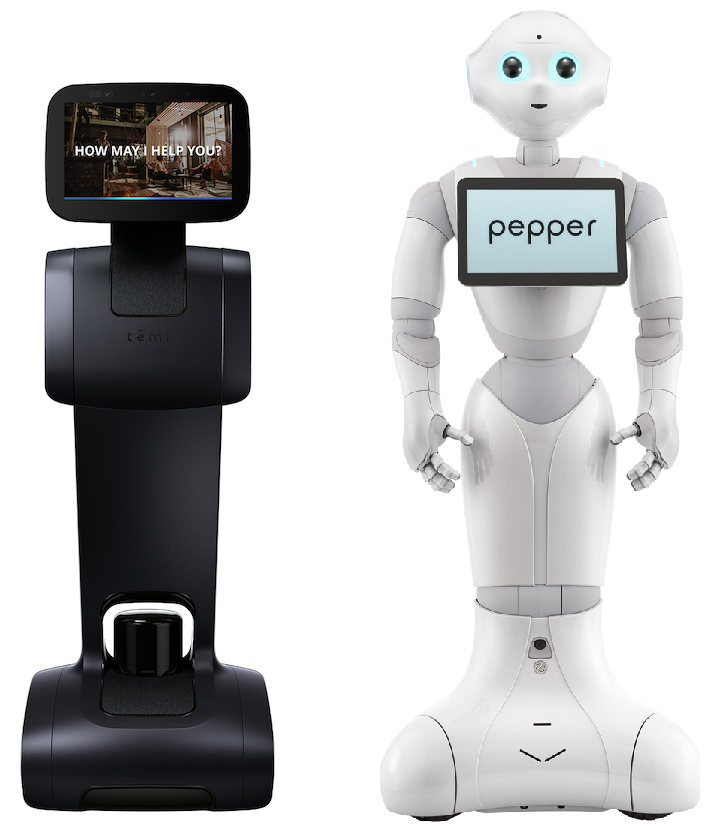}
    \caption{An image of the Temi and Pepper robots (not to scale) which provide a HRI interface for the customer service chatbot models.}
    \label{fig:temi-robot}
\end{figure}

\begin{figure}
    \centering
    \includegraphics[scale=0.19]{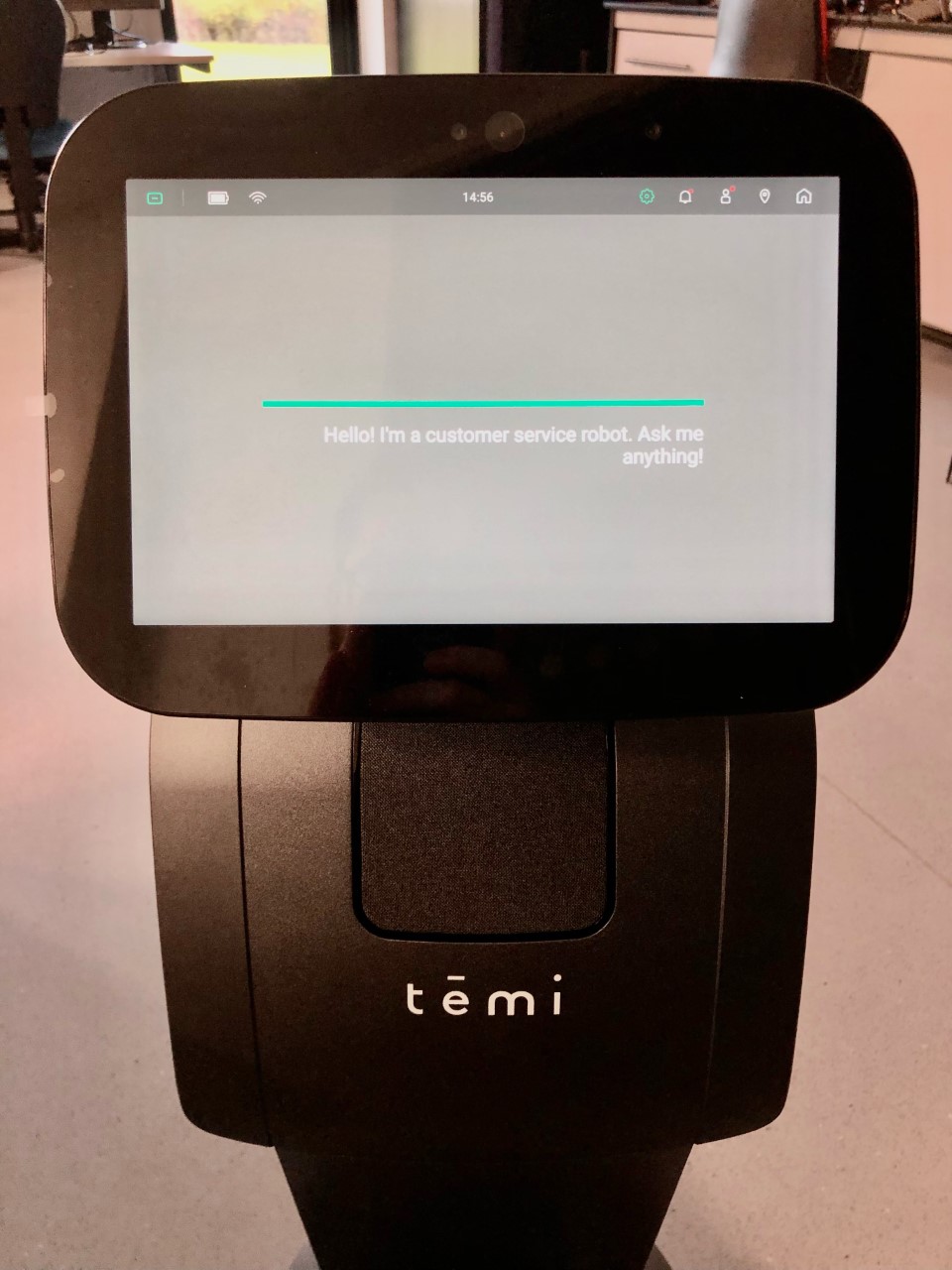}
    \caption{The user interface on Temi which allows for the user to speak. Once text has been extracted from audio, the user's message is inferred and responded to by the transformer-based chatbot. }
    \label{fig:temi_ui}
\end{figure}

This section details implementation of the chatbots to the Temi and Pepper robots, for purposes of both feasibility and proof-of-concept. During implementation, observations of drawbacks were made (e.g. with available consumer APIs), and solutions are presented. An image of Temi and Pepper can be seen in Figure \ref{fig:temi-robot}, with an image of the chatbot user interface in Figure \ref{fig:temi_ui}. 

\begin{table}[]
\caption{Time taken for the Temi robot to speak the first ten Harvard sentences.}
\label{tab:temi-wpm-table}
\begin{tabular}{@{}llll@{}}
\toprule
\textbf{Harvard Sentence} & \textbf{Number of words} & \textbf{Time Taken (s)} & \textbf{WPM} \\ \midrule
1                         & 8                        & 2.8                     & 171.43       \\
2                         & 8                        & 2.93                    & 163.82       \\
3                         & 9                        & 2.91                    & 185.57       \\
4                         & 9                        & 3.16                    & 170.89       \\
5                         & 7                        & 3.31                    & 126.89       \\
6                         & 7                        & 2.76                    & 152.17       \\
7                         & 8                        & 2.94                    & 163.27       \\
8                         & 8                        & 3.65                    & 131.51       \\
9                         & 7                        & 3.32                    & 126.51       \\
10                        & 8                        & 3.51                    & 136.75       \\ \midrule
\textit{\textbf{Average}} & 7.9                      & 3.129                   & 152.88       \\ \bottomrule
\end{tabular}
\end{table}

It was observed during the Temi implementation that, even when an await command is used, text-to-speech is not awaited until completion. This causes only the final speech command to be executed if there are multiple. Exploration discovered that this was due to a finished signal being sent upon communication of the command from a computer to the robotic device. To remedy this problem, the speed at which Temi speaks was measured in Table \ref{tab:temi-wpm-table}, where an average speaking speed of 152.88 words per minute was found. This provided a pointer to a sleep command parameter, so the text-to-speech transition was properly expected, simply calculated as $wait_{sec} =\frac{words}{WPM}\times60$. Although the average is used, future work should concern further exploration of synthetic speech speed. 
\begin{figure}
    \centering
    \includegraphics[scale=0.90]{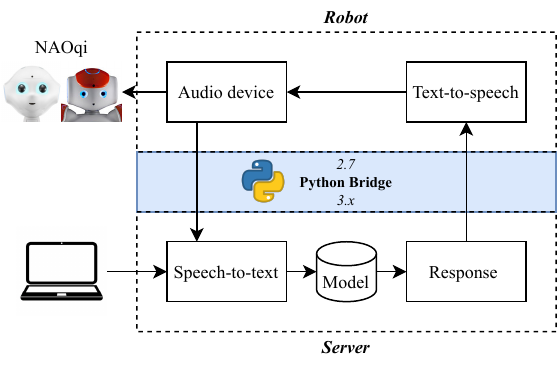}
    \caption{An implementation strategy of server-sided speech-to-text and a Python bridge to enable interfacing with the NAOqi SDK.}
    \label{fig:naoqi-diagram}
\end{figure}
This problem did not occur during the implementation on the Pepper robot, although other issues were observed. The issues were twofold; i) native speech recognition on the Softbank robots is limited to the recognition of keywords rather than speech-to-text, and ii) NAOqi is only available as a Python 2.7 implementation, and direct conversion to Python 3.x is not possible. Two solutions are proposed in the strategy detailed in Figure \ref{fig:naoqi-diagram}; speech recognition must instead be performed on the server side with text being processed and responses passed to the robot, and a bridge is instantiated to communicate between Python versions. TensorFlow is used to produce a prediction on the input data in Python 3.x (3.7 for this study), then operating in Python 2.7 to connect to the robot over a network and send the command across.

\section{Future Work and Conclusion}
\label{sec:futurework}
The marginal differences observed when tuning topology, as well as the results found in \cite{michel2019sixteen}, suggest that future models could be made less computationally complex while remaining accurate by way of pruning towards the most useful attention heads. Future work could explore pruning as a method to improve compatibility with weaker robotic hardware as well as the response time of the chatbot. Within the current version of TensorFlow (2.6.0), the tested metrics for sparse categorical learning are cross-entropy loss, accuracy, and top-k accuracy; this study focused on the analysis of these metrics with $k=5$ and $k=10$. If other metrics are tested and implemented in the future, then precision, recall, and F1 score etc. should be measured and compared as well. Alternatively, with a large amount of RAM required, future experiments could one-hot encode the model outputs to sparse matrices (rather than integer labels) and utilise categorical cross-entropy, enabling the aforementioned metrics. Given more computational resources, future work could also concern the transfer of knowledge from all-to-one domains rather than the one-to-one transfer learning experiments that were performed in this study. 

Several issues facing the feasibility of implementation were encountered during the latter half of these studies, and Section \ref{subsec:implementation} proposed solutions to overcome them, ultimately providing strategies to chatbot implementation on physical robots. The largest issues faced were due to the possibilities of the NAOqi SDK, with the main drawback in particular being that the SDK is unable to perform speech-to-text, rather, certain keywords are recognised instead. The solution to perform speech-to-text on another device allowed for a working implementation, and future work must therefore concern the possibility of embedded devices on the robots for more distant HRI to take place. Another issue faced was the lack of Python 3.x support for the NAOqi SDK, which is incompatible with recent implementations of TensorFlow. A Python bridge overcame this issue by communicating direct commands from a server which performed inference on the model. 

To finally conclude, this study has shown that knowledge transfer is possible between chatbots of several domains in order to improve the ability of autonomous customer support. In the majority of cases, several machine learning metrics were improved when knowledge was transferred from at least one other domain. During exploration of the models, it was seen that, in contrast to the more static nature of non-human support, relatively natural communication took place. Examples of this included the chatbots seemingly empathising with angrier customers and those faced with particularly difficult issues, as well as a change in tone of voice given the user's input, and also nuanced behaviours such as responding `hi' to customers who had begun their message with the same. The results found by the experiments in this article are promising, providing a method to enable better customer support chatbots in the future by sharing knowledge from other domains. 

\ifCLASSOPTIONcaptionsoff
  \newpage
\fi

\bibliographystyle{IEEEtran}
\bibliography{bibliography}

\begin{thebibliography}{10}
\providecommand{\url}[1]{#1}
\csname url@samestyle\endcsname
\providecommand{\newblock}{\relax}
\providecommand{\bibinfo}[2]{#2}
\providecommand{\BIBentrySTDinterwordspacing}{\spaceskip=0pt\relax}
\providecommand{\BIBentryALTinterwordstretchfactor}{4}
\providecommand{\BIBentryALTinterwordspacing}{\spaceskip=\fontdimen2\font plus
\BIBentryALTinterwordstretchfactor\fontdimen3\font minus
  \fontdimen4\font\relax}
\providecommand{\BIBforeignlanguage}[2]{{%
\expandafter\ifx\csname l@#1\endcsname\relax
\typeout{** WARNING: IEEEtran.bst: No hyphenation pattern has been}%
\typeout{** loaded for the language `#1'. Using the pattern for}%
\typeout{** the default language instead.}%
\else
\language=\csname l@#1\endcsname
\fi
#2}}
\providecommand{\BIBdecl}{\relax}
\BIBdecl

\bibitem{esmaeilian2016evolution}
B.~Esmaeilian, S.~Behdad, and B.~Wang, ``The evolution and future of
  manufacturing: A review,'' \emph{Journal of Manufacturing Systems}, vol.~39,
  pp. 79--100, 2016.

\bibitem{michalos2010automotive}
G.~Michalos, S.~Makris, N.~Papakostas, D.~Mourtzis, and G.~Chryssolouris,
  ``Automotive assembly technologies review: challenges and outlook for a
  flexible and adaptive approach,'' \emph{CIRP Journal of Manufacturing Science
  and Technology}, vol.~2, no.~2, pp. 81--91, 2010.

\bibitem{lunghi2019multimodal}
G.~Lunghi, R.~Marin, M.~Di~Castro, A.~Masi, and P.~J. Sanz, ``Multimodal
  human-robot interface for accessible remote robotic interventions in
  hazardous environments,'' \emph{IEEE Access}, vol.~7, pp. 127\,290--127\,319,
  2019.

\bibitem{wallop_2021}
\BIBentryALTinterwordspacing
H.~Wallop, ``On hold: How long it takes to speak to a human at major
  organisations,'' Jun 2021. [Online]. Available:
  \url{https://www.thisismoney.co.uk/money/news/article-9737653/On-hold-long-takes-speak-human-major-organisations.html}
\BIBentrySTDinterwordspacing

\bibitem{bygballe2012managing}
L.~E. Bygballe, E.~B{\o}, and S.~E. Gr{\o}nland, ``Managing international
  supply: The balance between total costs and customer service,''
  \emph{Industrial Marketing Management}, vol.~41, no.~3, pp. 394--401, 2012.

\bibitem{potter1994new}
J.~Potter-Brotman, ``The new role of service in customer retention,''
  \emph{Managing Service Quality: An International Journal}, 1994.

\bibitem{lokman2018modern}
A.~S. Lokman and M.~A. Ameedeen, ``Modern chatbot systems: A technical
  review,'' in \emph{Proceedings of the future technologies conference}.\hskip
  1em plus 0.5em minus 0.4em\relax Springer, 2018, pp. 1012--1023.

\bibitem{turing1950computing}
A.~M. Turing and J.~Haugeland, \emph{Computing machinery and
  intelligence}.\hskip 1em plus 0.5em minus 0.4em\relax MIT Press Cambridge,
  MA, 1950.

\bibitem{floridi2009turing}
L.~Floridi, M.~Taddeo, and M.~Turilli, ``Turing’s imitation game: still an
  impossible challenge for all machines and some judges----an evaluation of the
  2008 loebner contest,'' \emph{Minds and Machines}, vol.~19, no.~1, pp.
  145--150, 2009.

\bibitem{cui2017superagent}
L.~Cui, S.~Huang, F.~Wei, C.~Tan, C.~Duan, and M.~Zhou, ``Superagent: A
  customer service chatbot for e-commerce websites,'' in \emph{Proceedings of
  ACL 2017, System Demonstrations}, 2017, pp. 97--102.

\bibitem{nuruzzaman2018survey}
M.~Nuruzzaman and O.~K. Hussain, ``A survey on chatbot implementation in
  customer service industry through deep neural networks,'' in \emph{2018 IEEE
  15th International Conference on e-Business Engineering (ICEBE)}.\hskip 1em
  plus 0.5em minus 0.4em\relax IEEE, 2018, pp. 54--61.

\bibitem{xu2017new}
A.~Xu, Z.~Liu, Y.~Guo, V.~Sinha, and R.~Akkiraju, ``A new chatbot for customer
  service on social media,'' in \emph{Proceedings of the 2017 CHI conference on
  human factors in computing systems}, 2017, pp. 3506--3510.

\bibitem{ranoliya2017chatbot}
B.~R. Ranoliya, N.~Raghuwanshi, and S.~Singh, ``Chatbot for university related
  faqs,'' in \emph{2017 International Conference on Advances in Computing,
  Communications and Informatics (ICACCI)}.\hskip 1em plus 0.5em minus
  0.4em\relax IEEE, 2017, pp. 1525--1530.

\bibitem{feine2019measuring}
J.~Feine, S.~Morana, and U.~Gnewuch, ``Measuring service encounter satisfaction
  with customer service chatbots using sentiment analysis,'' 2019.

\bibitem{tran2021exploring}
A.~D. Tran, J.~I. Pallant, and L.~W. Johnson, ``Exploring the impact of
  chatbots on consumer sentiment and expectations in retail,'' \emph{Journal of
  Retailing and Consumer Services}, vol.~63, p. 102718, 2021.

\bibitem{bird2019high}
J.~J. Bird, A.~Ek{\'a}rt, C.~D. Buckingham, and D.~R. Faria, ``High resolution
  sentiment analysis by ensemble classification,'' in \emph{Intelligent
  Computing-Proceedings of the Computing Conference}.\hskip 1em plus 0.5em
  minus 0.4em\relax Springer, 2019, pp. 593--606.

\bibitem{jia2021chinese}
K.~Jia, ``Chinese sentiment classification based on word2vec and vector
  arithmetic in human--robot conversation,'' \emph{Computers \& Electrical
  Engineering}, vol.~95, p. 107423, 2021.

\bibitem{ricciardelli2019self}
E.~Ricciardelli and D.~Biswas, ``Self-improving chatbots based on reinforcement
  learning,'' in \emph{4th Multidisciplinary Conference on Reinforcement
  Learning and Decision Making}, 2019.

\bibitem{cuayahuitl2019ensemble}
H.~Cuay{\'a}huitl, D.~Lee, S.~Ryu, Y.~Cho, S.~Choi, S.~Indurthi, S.~Yu,
  H.~Choi, I.~Hwang, and J.~Kim, ``Ensemble-based deep reinforcement learning
  for chatbots,'' \emph{Neurocomputing}, vol. 366, pp. 118--130, 2019.

\bibitem{bali2019diabot}
M.~Bali, S.~Mohanty, S.~Chatterjee, M.~Sarma, and R.~Puravankara, ``Diabot: a
  predictive medical chatbot using ensemble learning,'' \emph{Int. J. of Recent
  Technol. and Eng.}, pp. 2277--3878, 2019.

\bibitem{harilal2020caro}
N.~Harilal, R.~Shah, S.~Sharma, and V.~Bhutani, ``Caro: an empathetic health
  conversational chatbot for people with major depression,'' in
  \emph{Proceedings of the 7th ACM IKDD CoDS and 25th COMAD}, 2020, pp.
  349--350.

\bibitem{mondal2018chatbot}
A.~Mondal, M.~Dey, D.~Das, S.~Nagpal, and K.~Garda, ``Chatbot: An automated
  conversation system for the educational domain,'' in \emph{2018 International
  Joint Symposium on Artificial Intelligence and Natural Language Processing
  (iSAI-NLP)}.\hskip 1em plus 0.5em minus 0.4em\relax IEEE, 2018, pp. 1--5.

\bibitem{bird2021chatbot}
J.~J. Bird, A.~Ek{\'a}rt, and D.~R. Faria, ``Chatbot interaction with
  artificial intelligence: human data augmentation with t5 and language
  transformer ensemble for text classification,'' \emph{Journal of Ambient
  Intelligence and Humanized Computing}, pp. 1--16, 2021.

\bibitem{ilievski2018goal}
V.~Ilievski, C.~Musat, A.~Hossmann, and M.~Baeriswyl, ``Goal-oriented chatbot
  dialog management bootstrapping with transfer learning,'' in \emph{IJCAI},
  2018.

\bibitem{hatua2019goal}
A.~Hatua, T.~T. Nguyen, and A.~H. Sung, ``Goal-oriented conversational system
  using transfer learning and attention mechanism,'' in \emph{2019 IEEE 10th
  Annual Ubiquitous Computing, Electronics \& Mobile Communication Conference
  (UEMCON)}.\hskip 1em plus 0.5em minus 0.4em\relax IEEE, 2019, pp. 0099--0104.

\bibitem{yu2018modelling}
J.~Yu, M.~Qiu, J.~Jiang, J.~Huang, S.~Song, W.~Chu, and H.~Chen, ``Modelling
  domain relationships for transfer learning on retrieval-based question
  answering systems in e-commerce,'' in \emph{Proceedings of the Eleventh ACM
  International Conference on Web Search and Data Mining}, 2018, pp. 682--690.

\bibitem{yam2020robots}
K.~C. Yam, Y.~E. Bigman, P.~M. Tang, R.~Ilies, D.~De~Cremer, H.~Soh, and
  K.~Gray, ``Robots at work: People prefer—and forgive—service robots with
  perceived feelings.'' \emph{Journal of Applied Psychology}, 2020.

\bibitem{mori1970bukimi}
M.~Mori, ``Bukimi no tani [the uncanny valley],'' \emph{Energy}, vol.~7, pp.
  33--35, 1970.

\bibitem{murphy2017service}
J.~Murphy, U.~Gretzel, and C.~Hofacker, ``Service robots in hospitality and
  tourism: investigating anthropomorphism,'' in \emph{15th APacCHRIE
  conference}, vol.~31, 2017.

\bibitem{wirtz2018brave}
J.~Wirtz, P.~G. Patterson, W.~H. Kunz, T.~Gruber, V.~N. Lu, S.~Paluch, and
  A.~Martins, ``Brave new world: service robots in the frontline,''
  \emph{Journal of Service Management}, 2018.

\bibitem{tuomi2021applications}
A.~Tuomi, I.~P. Tussyadiah, and J.~Stienmetz, ``Applications and implications
  of service robots in hospitality,'' \emph{Cornell Hospitality Quarterly},
  vol.~62, no.~2, pp. 232--247, 2021.

\bibitem{vaswani2017attention}
A.~Vaswani, N.~Shazeer, N.~Parmar, J.~Uszkoreit, L.~Jones, A.~N. Gomez,
  {\L}.~Kaiser, and I.~Polosukhin, ``Attention is all you need,'' in
  \emph{Advances in neural information processing systems}, 2017, pp.
  5998--6008.

\bibitem{devlin2018open}
J.~Devlin and M.-W. Chang, ``Open sourcing bert: State-of-the-art pre-training
  for natural language processing,'' \emph{Google AI Blog. Weblog.[Online]
  Available from: https://ai. googleblog.
  com/2018/11/open-sourcing-bertstate-of-art-pre. html [Accessed 4 December
  2019]}, 2018.

\bibitem{radford2019language}
A.~Radford, J.~Wu, R.~Child, D.~Luan, D.~Amodei, and I.~Sutskever, ``Language
  models are unsupervised multitask learners,'' 2019.

\bibitem{shao2019transformer}
T.~Shao, Y.~Guo, H.~Chen, and Z.~Hao, ``Transformer-based neural network for
  answer selection in question answering,'' \emph{IEEE Access}, vol.~7, pp.
  26\,146--26\,156, 2019.

\bibitem{lukovnikov2019pretrained}
D.~Lukovnikov, A.~Fischer, and J.~Lehmann, ``Pretrained transformers for simple
  question answering over knowledge graphs,'' in \emph{International Semantic
  Web Conference}.\hskip 1em plus 0.5em minus 0.4em\relax Springer, 2019, pp.
  470--486.

\bibitem{vector_2017}
\BIBentryALTinterwordspacing
{Thought Vector}, ``Dataset: Customer support on twitter.'' [Online].
  Available:
  \url{https://www.kaggle.com/thoughtvector/customer-support-on-twitter}
\BIBentrySTDinterwordspacing

\bibitem{gabriel_2016}
\BIBentryALTinterwordspacing
R.~J. Gabriel, ``Github repository: Robertjgabriel/google-profanity-words,''
  2016. [Online]. Available:
  \url{https://github.com/RobertJGabriel/Google-profanity-words}
\BIBentrySTDinterwordspacing

\bibitem{michel2019sixteen}
P.~Michel, O.~Levy, and G.~Neubig, ``Are sixteen heads really better than
  one?'' \emph{Advances in Neural Information Processing Systems}, vol.~32, pp.
  14\,014--14\,024, 2019.

\end{thebibliography}

\end{document}